\documentclass[trackchanges,twocolumn]{aastex701}

\usepackage[normalem]{ulem} 

\begin{document}

\title{Vector Magnetic Field associated with an Active Region Filament Observed by SUNRISE III/SCIP in the Ca II 8542 {\AA} Line}

\author[0000-0002-1043-9944]{Takuma Matsumoto}
\affiliation{Center for Integrated Data Science, Institute for Space-Earth Environmental Research, Nagoya University, Furocho, Chikusa-ku, Nagoya, Aichi, 464-8601, Japan}
\affiliation{National Astronomical Observatory of Japan, 2-21-1 Osawa, Mitaka, Tokyo, 181-8588, Japan}
\email{takuma.matsumoto@gmail.com}

\author[0000-0002-5054-8782]{Yukio Katsukawa}
\affiliation{National Astronomical Observatory of Japan, 2-21-1 Osawa, Mitaka, Tokyo, 181-8588, Japan}
\affiliation{Department of Astronomical Science, School of Physical Sciences, The Graduate University for Advanced Studies, Sokendai, 2-21-1 Osawa, Mitaka, Tokyo, 181-8588, Japan}
\email{yukio.katsukawa@nao.ac.jp}

\author[0000-0001-5616-2808]{Masahito Kubo}
\affiliation{National Astronomical Observatory of Japan, 2-21-1 Osawa, Mitaka, Tokyo, 181-8588, Japan}
\email{masahito.kubo@nao.ac.jp}

\author[0000-0001-7452-0656]{Yusuke Kawabata}
\affiliation{National Astronomical Observatory of Japan, 2-21-1 Osawa, Mitaka, Tokyo, 181-8588, Japan}
\email{yusuke.kawabata@nao.ac.jp}

\author[0000-0002-7044-6281]{Takayoshi Oba}
\affiliation{Advanced Research Center for Space Science and Technology, Institute of Science and Engineering, Kanazawa University, Kakuma-machi, Kanazawa, Ishikawa 920-1192, Japan}
\affiliation{Max Planck Institute for Solar System Research, Justus-von-Liebig-Weg 3, 37077 G\"{o}ttingen, Germany}
\email{oba-takayoshi@staff.kanazawa-u.ac.jp}

\author[0000-0002-4669-5376]{Ryohtaroh T. Ishikawa}
\affiliation{National Institute for Fusion Science, 322-6 Oroshi-chi, Toki, Gifu 509-5292, Japan}
\affiliation{Fusion Science Program, Graduate Institute for Advanced Studies, The Graduate University for Advanced Studies, SOKENDAI, 322-6 Oroshi-chi, Toki, Gifu 5095292, Japan}
\email{ishikawa.ryohtaro@nifs.ac.jp}

\author[0000-0001-6793-8528]{Yoshihiro Naito}
\affiliation{Department of Astronomical Science, School of Physical Sciences, The Graduate University for Advanced Studies, Sokendai, 2-21-1 Osawa, Mitaka, Tokyo, 181-8588, Japan}
\affiliation{National Astronomical Observatory of Japan, 2-21-1 Osawa, Mitaka, Tokyo, 181-8588, Japan}
\email{yoshihiro.naito@grad.nao.ac.jp}

\author[0000-0001-5686-3081]{Hirohisa Hara}
\affiliation{National Astronomical Observatory of Japan, 2-21-1 Osawa, Mitaka, Tokyo, 181-8588, Japan}
\affiliation{Department of Astronomical Science, School of Physical Sciences, The Graduate University for Advanced Studies, Sokendai, 2-21-1 Osawa, Mitaka, Tokyo, 181-8588, Japan}
\email{hirohisa.hara@nao.ac.jp}

\author[0000-0003-4764-6856]{Toshifumi Shimizu}
\affiliation{Institute of Space and Astronautical Science, Japan Aerospace Exploration Agency, 3-1-1 Chuo-ku, Sagamihara, Kanagawa, 252-5210, Japan}
\affiliation{Department of Earth and Planetary Science, The University of Tokyo, 7-3-1 Hongo, Bunkyo-ku, Tokyo 113-0033, Japan}
\email{shimizu.toshifumi@jaxa.jp}

\author[0009-0005-9709-8431]{Fumihiro Uraguchi}
\affiliation{National Astronomical Observatory of Japan, 2-21-1 Osawa, Mitaka, Tokyo, 181-8588, Japan}
\email{fumihiro.uraguchi@nao.ac.jp}

\author[0000-0002-8342-8314]{Toshihiro Tsuzuki}
\affiliation{National Astronomical Observatory of Japan, 2-21-1 Osawa, Mitaka, Tokyo, 181-8588, Japan}
\email{toshihiro.tsuzuki@nao.ac.jp}

\author{Kazuya Shinoda}
\affiliation{National Astronomical Observatory of Japan, 2-21-1 Osawa, Mitaka, Tokyo, 181-8588, Japan}
\email{shinoda.kazuya@nao.ac.jp}

\author{Tomonori Tamura}
\affiliation{National Astronomical Observatory of Japan, 2-21-1 Osawa, Mitaka, Tokyo, 181-8588, Japan}
\email{tomonori.tamura@nao.ac.jp}

\author[0000-0003-4452-858X]{Yoshinori Suematsu}
\affiliation{National Astronomical Observatory of Japan, 2-21-1 Osawa, Mitaka, Tokyo, 181-8588, Japan}
\email{yoshinori.suematsu@nao.ac.jp}

\author[0000-0001-5518-8782]{Carlos Quintero Noda}
\affiliation{Instituto de Astrofísica de Canarias, E-38205 La Laguna, Tenerife, Spain}
\affiliation{Departamento de Astrof\'{i}sica, Univ. de La Laguna, E-38205 La Laguna, Tenerife, Spain}
\email{carlos.quintero@iac.es}

\author[0000-0002-3418-8449]{Sami~K.~Solanki} \affiliation{Max-Planck-Institut für Sonnensystemforschung, Justus-von-Liebig-Weg 3, 37077 Göttingen, Germany}\email{solanki@mps.mpg.de}

\author[0000-0003-1459-7074]{Andreas~Lagg} \affiliation{Max-Planck-Institut für Sonnensystemforschung, Justus-von-Liebig-Weg 3, 37077 Göttingen, Germany}\email{lagg@mps.mpg.de}
\author[0000-0002-9972-9840]{Achim~Gandorfer} \affiliation{Max-Planck-Institut für Sonnensystemforschung, Justus-von-Liebig-Weg 3, 37077 Göttingen, Germany}\email{gandorfer@mps.mpg.de}
\author[0000-0002-3387-026X]{Jose~Carlos~del~Toro~Iniesta} \affiliation{Instituto de Astrofísica de Andalucía, CSIC, Glorieta de la Astronomía s/n, 18008 Granada, Spain}\affiliation{Spanish Space Solar Physics Consortium}\email{jti@iaa.es}
\author[0000-0002-0787-8954]{Pietro~Bernasconi} \affiliation{Johns Hopkins University Applied Physics Laboratory, 11100 Johns Hopkins Road, Laurel, Maryland, USA}\email{pietro.bernasconi@jhuapl.edu}
\author[sname='Berkefeld']{Thomas~Berkefeld} \affiliation{Institut für Sonnenphysik (KIS), Georges-Köhler-Allee 401a, 79110 Freiburg, Germany}\email{thomas.berkefeld@leibniz-kis.de}
\author[0009-0009-4425-599X]{Alex~Feller} \affiliation{Max-Planck-Institut für Sonnensystemforschung, Justus-von-Liebig-Weg 3, 37077 Göttingen, Germany}\email{feller@mps.mpg.de}
\author[0000-0001-6317-4380]{Tino~L.~Riethmüller} \affiliation{Max-Planck-Institut für Sonnensystemforschung, Justus-von-Liebig-Weg 3, 37077 Göttingen, Germany}\email{riethmueller@mps.mpg.de}

\author[0000-0001-9228-3412]{Alberto~Álvarez-Herrero} \affiliation{Instituto Nacional de T\'ecnica Aeroespacial (INTA), Ctra. de Ajalvir, km. 4, E-28850 Torrejón de Ardoz, Spain}\affiliation{Spanish Space Solar Physics Consortium}\email{alvareza@inta.es}
\author[0000-0003-3490-6532]{H.~N.~Smitha} \affiliation{Max-Planck-Institut für Sonnensystemforschung, Justus-von-Liebig-Weg 3, 37077 Göttingen, Germany}\email{narayanamurthy@mps.mpg.de}
\author[0000-0001-8829-1938]{David~Orozco~Suárez} \affiliation{Instituto de Astrofísica de Andalucía, CSIC, Glorieta de la Astronomía s/n, 18008 Granada, Spain}\affiliation{Spanish Space Solar Physics Consortium}\email{orozco@iaa.es}
\author[sname='Grauf']{Bianca~Grauf} \affiliation{Max-Planck-Institut für Sonnensystemforschung, Justus-von-Liebig-Weg 3, 37077 Göttingen, Germany}\email{grauf@mps.mpg.de}
\author[sname='Carpenter']{Michael~Carpenter} \affiliation{Johns Hopkins University Applied Physics Laboratory, 11100 Johns Hopkins Road, Laurel, Maryland, USA}\email{michael.carpenter@jhuapl.edu}
\author[sname='Bell']{Alexander~Bell} \affiliation{Institut für Sonnenphysik (KIS), Georges-Köhler-Allee 401a, 79110 Freiburg, Germany}\email{albe@leibniz-kis.de}
\author[0000-0001-7764-6895]{Valentín~Martínez~Pillet} \affiliation{Instituto de Astrofísica de Canarias, Vía Láctea, s/n, E-38205 La Laguna, Spain}\affiliation{Spanish Space Solar Physics Consortium}\email{vmpillet@iac.es}

\author[0000-0002-7318-3536]{Francisco~Javier~Bailén} \affiliation{Instituto de Astrofísica de Andalucía, CSIC, Glorieta de la Astronomía s/n, 18008 Granada, Spain}\affiliation{Spanish Space Solar Physics Consortium}\email{fbailen@iaa.es}
\author[0000-0002-2055-441X]{Julian~Blanco~Rodríguez} \affiliation{Universitat de Valencia Catedrático José Beltrán 2, E-46980 Paterna-Valencia, Spain}\affiliation{Spanish Space Solar Physics Consortium}\email{julian.blanco@uv.es}
\author[0000-0003-4319-2009]{Juan~Sebastián~Castellanos~Durán} \affiliation{Max-Planck-Institut für Sonnensystemforschung, Justus-von-Liebig-Weg 3, 37077 Göttingen, Germany}\email{castellanos@mps.mpg.de}
\author[0009-0002-6808-5154]{Edvarda~Harnes} \affiliation{Max-Planck-Institut für Sonnensystemforschung, Justus-von-Liebig-Weg 3, 37077 Göttingen, Germany}\email{harnes@mps.mpg.de}
\author[0000-0001-6029-7529]{Johannes~Hölken} \affiliation{Max-Planck-Institut für Sonnensystemforschung, Justus-von-Liebig-Weg 3, 37077 Göttingen, Germany}\email{hoelken@mps.mpg.de}
\author[0000-0003-1409-1145]{Francisco~A.~Iglesias} \affiliation{Max-Planck-Institut für Sonnensystemforschung, Justus-von-Liebig-Weg 3, 37077 Göttingen, Germany}\affiliation{Grupo de Estudios en Heliofísica de Mendoza, CONICET, Universidad de Mendoza, Boulogne sur Mer 683, 5500 Mendoza, Argentina}\email{iglesias@mps.mpg.de}
\author[0000-0003-0175-6232]{Azaymi~L.~Siu-Tapia} \affiliation{Instituto de Astrofísica de Andalucía, CSIC, Glorieta de la Astronomía s/n, 18008 Granada, Spain}\affiliation{Spanish Space Solar Physics Consortium}\email{siu@iaa.es}
\author[0000-0003-1483-4535]{Hanna~Strecker} \affiliation{Instituto de Astrofísica de Andalucía, CSIC, Glorieta de la Astronomía s/n, 18008 Granada, Spain}\affiliation{Spanish Space Solar Physics Consortium}\email{streckerh@iaa.es}
\author[0000-0003-1971-5551]{Dušan~Vukadinović} \affiliation{Institut für Physik, Universität Graz, Universitätsplatz 5, 8010 Graz, Austria}\affiliation{Max-Planck-Institut für Sonnensystemforschung, Justus-von-Liebig-Weg 3, 37077 Göttingen, Germany}\email{dusan.vukadinovic@uni-graz.at}

\author[0000-0001-7094-518X]{Pablo~Santamarina~Guerrero} \affiliation{Instituto de Astrofísica de Andalucía, CSIC, Glorieta de la Astronomía s/n, 18008 Granada, Spain}\affiliation{Spanish Space Solar Physics Consortium}\email{psanta@iaa.es}

\author[0000-0003-2409-3742]{Nour~E.~Raouafi} \affiliation{Johns Hopkins University Applied Physics Laboratory, 11100 Johns Hopkins Road, Laurel, Maryland, USA}\email{Nour.Raouafi@jhuapl.edu}

\begin{abstract}
We report high-spatial-resolution spectropolarimetric observations spatially associated with a solar filament, obtained with the \textsc{Sunrise} Chromospheric Infrared spectro-Polarimeter (SCIP) onboard the \textsc{Sunrise~iii} balloon-borne solar observatory on 15 July 2024.
The observed filament was located near the solar disk center, adjacent to an active region, and remained quiescent for at least two hours during the observing period.
SCIP recorded full Stokes profiles in the \ion{Ca}{2} 8542 {\AA} line, revealing clear signatures of linear polarization produced by the transverse Zeeman effect.
The detected linear polarization signals within the filament region exceeded the 2$\sigma$ noise level and exhibited a characteristic Zeeman double-lobe spectral shape that distinguishes them from polarization due to scattering.
The magnetic field strength derived using the weak field approximation is approximately $-80$ G along the line of sight and 300–500 G in the transverse direction.
These values likely reflect the magnetic properties of the filament and its supporting chromospheric environment.
The orientation of the magnetic field vector is nearly parallel to the filament axis in its northeastern portion, while the southeastern part of the filament extends outside the field of view.
To our knowledge, this is the first unambiguous detection of linear polarization associated with a solar filament with the \ion{Ca}{2} 8542 {\AA} line.
Our results open a new diagnostic window on the vector magnetic structure of solar filaments in the lower chromosphere, complementing existing \ion{He}{1} based diagnostics that probe the upper chromosphere.
\end{abstract}


\section{Introduction} \label{sec:intro}

Solar filaments are dense, cool structures suspended in the hot corona, and their stability and eruption are governed by the underlying magnetic field configuration \citep{2018LRSP...15....7G}. 
Despite their importance in triggering coronal mass ejections, direct measurements of the magnetic field vector in filaments remain challenging due to their complex geometry, low signal strength in polarization, and the limited sensitivity of current instrumentation.

Early investigations of the magnetic field in solar filaments relied on various spectral lines, most notably \ion{He}{1} $D_3$, H$\alpha$, and H$\beta$ \citep{1983SoPh...83..135L, 1984A&A...131...33L,1994SoPh..154..231B}.
More recently, studies have shifted their focus toward the \ion{He}{1} 10830 {\AA} triplet to investigate the magnetic structure in both quiet-Sun filaments \citep{2014A&A...566A..46O,2015ApJ...802....3M,2020ApJ...892...75W,2023PASJ...75..660Y} and active-region filaments \citep{1995ApJ...441L..51P,2009A&A...501.1113K,2011A&A...526A..42S,2012ApJ...749..138X,2014A&A...561A..98S,2016ApJ...822...50D,2026PASJ...78..252H}.
The triplet requires EUV irradiation to form and is therefore sensitive to upper-chromospheric plasma \citep{1994IAUS..154...35A}.
It is affected by both the Hanle and the Zeeman effects over a wide range of magnetic field strengths up to $\sim$1000~G \citep{2007ApJ...655..642T}, which makes the interpretation of the polarization signals nontrivial.
Although the polarization amplitudes are relatively large, contamination from the surrounding, non-filament atmosphere has been pointed out in some cases \citep{2016ApJ...822...50D}.

The \ion{Ca}{2} 8542 {\AA} line, on the other hand, forms mainly in the lower chromosphere \citep{2009ApJ...694L.128L,2017MNRAS.470.1453Q}.
For magnetic fields stronger than about 100 G, the Zeeman effect dominates \citep{2016ApJ...826L..10S}, which simplifies the analysis.
However, the polarization signals are much weaker, and polarimetric observations of filaments in this line are scarce \citep{1991A&A...247..379W,2012SoPh..280...69H,2019A&A...623A.178D}.
To our knowledge, a clear detection of the transverse component in filaments using infrared \ion{Ca}{2} lines has not been reported so far; only upper limits on the transverse field strength have been established by the detection limits \citep{2019A&A...625A.128D}.

In this Letter, we present high-resolution spectropolarimetric observations associated with a solar filament using the \textsc{Sunrise} Chromospheric Infrared spectro-Polarimeter (SCIP; \citealt{2020SPIE11447E..0YK, 2026arXiv260317929K}) onboard the \textsc{Sunrise~iii} balloon-borne observatory \citep{2025SoPh..300...75K,2026arXiv260607989S}, which represents the third flight of the \textsc{Sunrise} mission series \citep{2010ApJ...723L.127S, 2011SoPh..268....1B, 2017ApJS..229....2S}.
We report significant linear polarization signals in the \ion{Ca}{2} 8542 {\AA} line associated with a filament, enabling us to infer the magnetic field vector using Zeeman-effect-based diagnostics.

\section{Observations} \label{sec:obs}

We report observations obtained with SCIP onboard the \textsc{Sunrise~iii} balloon-borne solar observatory. 
SCIP simultaneously performs full-Stokes spectropolarimetry in two near-infrared spectral bands around 770 nm and 850 nm. 
In this study, we focus on the data from the 850 nm band, which includes the chromospheric \ion{Ca}{2} 8542 {\AA} line (Land\'{e} factor, $g_{\rm eff} = 1.1)$ and the photospheric \ion{Fe}{1} 8468 {\AA} line ($g_{\rm eff}=2.5$), both of which are suitable for magnetic field diagnostics at their respective formation heights. 
Our target filament was observed on 15 July 2024, from 18:39:25 to 18:51:20 UT.
The field of view of SCIP is approximately 58\arcsec $\times$ 58\arcsec, with 1 second integration time per scan step.
The spatial sampling is 0.094{\arcsec} per pixel, and the spectral sampling is adjusted to 39.5 m{\AA} per pixel in the 850 nm band.
The observed region covers a plage area located southeast of NOAA Active Region 13744, which included a prominent active region filament situated near disk center at the time of observation.

\begin{figure}
	\includegraphics[width=9cm]{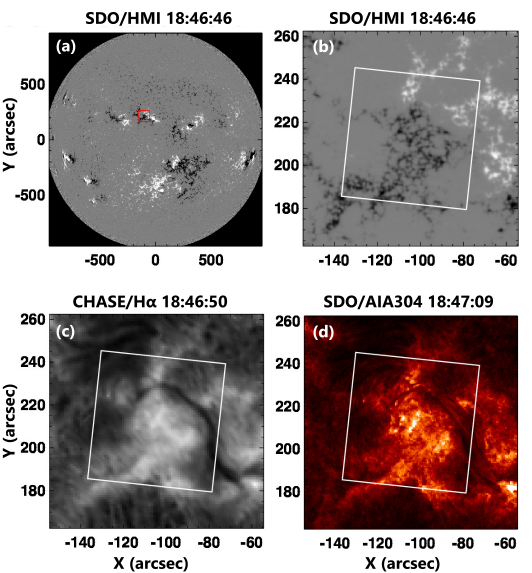}
    \caption{ 
    (a) Full-disk magnetogram from SDO/HMI with a red square indicating the FOV shown in the other panels.
    (b–d) Zoomed-in views of the region outlined in (a), showing (b) SDO/HMI magnetogram, (c) CHASE/H$\alpha$ line center, and (d) SDO/AIA 304 {\AA} observations.
    The white square in panels (b-d) indicates the FOV of the SCIP raster scan.
    }
    \label{fig:fov}
\end{figure}

Figure \ref{fig:fov}(a) presents a full-disk line-of-sight magnetogram from the Helioseismic and Magnetic Imager \citep[HMI;][]{2012SoPh..275..207S} onboard the Solar Dynamics Observatory \citep[SDO;][]{2012SoPh..275....3P}. 
The red square marks the region of interest, which is enlarged in panels (b)–(d). 
Panel (b) shows a close-up of the HMI magnetogram, while panels (c) and (d) display co-aligned images from the Chinese H$\alpha$ Solar Explorer \citep[CHASE;][]{2022SCPMA..6589602L} at H$\alpha$ line center, and the Atmospheric Imaging Assembly \citep[AIA;][]{2012SoPh..275...17L} at 304 {\AA}, respectively.
The white tilted rectangle overlaid in each panel indicates the FOV scanned by SCIP.
The SCIP slit was oriented approximately in the solar north–south direction, with the scanning motion proceeding from west to east. 
A dark filament, the main target of this study, is visible above the polarity inversion line in panel (b) and appears prominently in both the H$\alpha$ image and coronal image in panels (c) and (d).
The northeastern part of the filament is captured within the SCIP FOV, while the southwestern part lies outside the FOV (Fig \ref{fig:fov}c,d).

The standard SCIP data calibration pipeline was employed for dark-current subtraction, rolling-shutter correction, skew and wavelength calibration, flat-field correction, polarimetric correction, and cross-talk removal.
The estimated noise level in the calibrated Stokes $Q$, $U$, and $V$ signals was approximately $10^{-3}$ in units of the continuum intensity.

The orientation of the linear polarization signals was calibrated using simultaneous observations from the Hinode Spectro-Polarimeter \citep{2007SoPh..243....3K,2008SoPh..249..167T,2013SoPh..283..579L}.
We further verified the consistency of the calibration by confirming that the polarization azimuth of \ion{Ca}{2} 8542 {\AA} obtained from SCIP sunspot observations is nearly parallel to the radial structures of the surrounding penumbral filaments.
The high polarimetric sensitivity and spectral fidelity of SCIP, combined with the stability of the balloon platform and the on-board image stabilization system \citep{2025SoPh..300..112B,2026arXiv260207448B}, enable precise spectropolarimetric analysis of weak chromospheric magnetic fields.
The variation in the image rotation angle during the observation was approximately $1.7^\circ$, which is sufficiently small to have no significant impact on the current analysis.

\section{Results and Discussions} \label{sec:result}

\begin{figure*}
	\includegraphics[width=\linewidth]{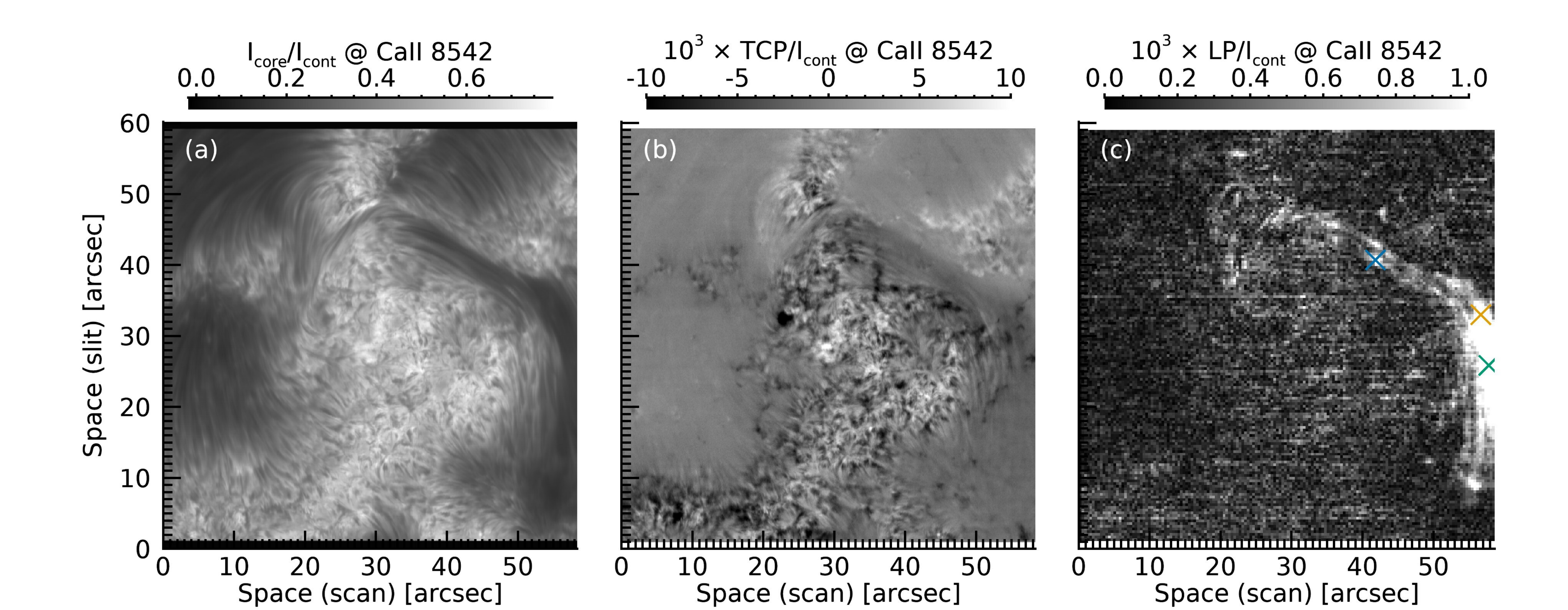}
    \caption{ 
      The scan map of SCIP for \ion{Ca}{2} 8542 {\AA} line indicating (a) line core intensity, (b) total circular polarization, and (c) linear polarization ($LP=\sqrt{Q^2 + U^2}$, averaged over $\pm 0.2$ {\AA}).
      Both the total and linear polarization signals are normalized by continuum intensity.
      The linear polarization map was binned over 4$\times$4 pixels to increase the S/N ratio.
    }
    \label{fig:scanmap}
\end{figure*}

Figure \ref{fig:scanmap} shows (a) the line-core intensity, (b) the total circular polarization ($TCP$), and (c) the linear polarization ($LP=\sqrt{Q^2+U^2}$, averaged over $\pm 0.2$ {\AA}) maps in the \ion{Ca}{2} 8542 {\AA} line.
The $TCP$ was calculated by subtracting Stokes $V$ in the red wing from Stokes $V$ in the blue wing \citep[e.g.][]{2013ApJ...768...69B}.
The circular and linear polarization signals were normalized by the continuum intensity level, which was defined as the median value of the full field of view.
The line-core image displays both fibrils and dark filamentary structures, with the central part of the field of view dominated by plage regions.
The filament consists of multiple fine threads that run nearly parallel to its axis, forming a coherent bundle of closely aligned structures.
The $TCP$ map shows consistently weak negative values within the filament location.
Remarkably, the $LP$ signals are clearly detected along the filament region, spatially coinciding with the absorption structures in panel (a) and basically lying above the polarity inversion line (Fig. \ref{fig:fov}b).
To increase the S/N ratio, the $Q$ and $U$ maps were spatially binned over 4$\times$4 pixels, which decrease the noise level down to $4\times 10^{-4}$ $I_{\rm cont}$ while reducing the spatial sampling to 0.34\arcsec.
The detection of significant $LP$ associated with a solar filament in the \ion{Ca}{2} 8542 {\AA} line represents a substantial advance, as previous studies could only place upper limits due to insufficient sensitivity or weak magnetic field \citep{2019A&A...623A.178D}. 
The $LP$ detections well exceed the $10^{-3}$ $I_{\rm cont}$ level with signal-to-noise ratios high enough ($>2\sigma$). 
We also detected $LP$ signals in the \ion{Ca}{2} 8498 {\AA} line, albeit systematically weaker, likely owing to its smaller effective Land\'{e} factor ($g_{\rm eff} = 1.07$) and slightly different formation properties; 
therefore, we focus on the \ion{Ca}{2} 8542 {\AA} results in this study.

\begin{figure}
	\includegraphics[width=9cm]{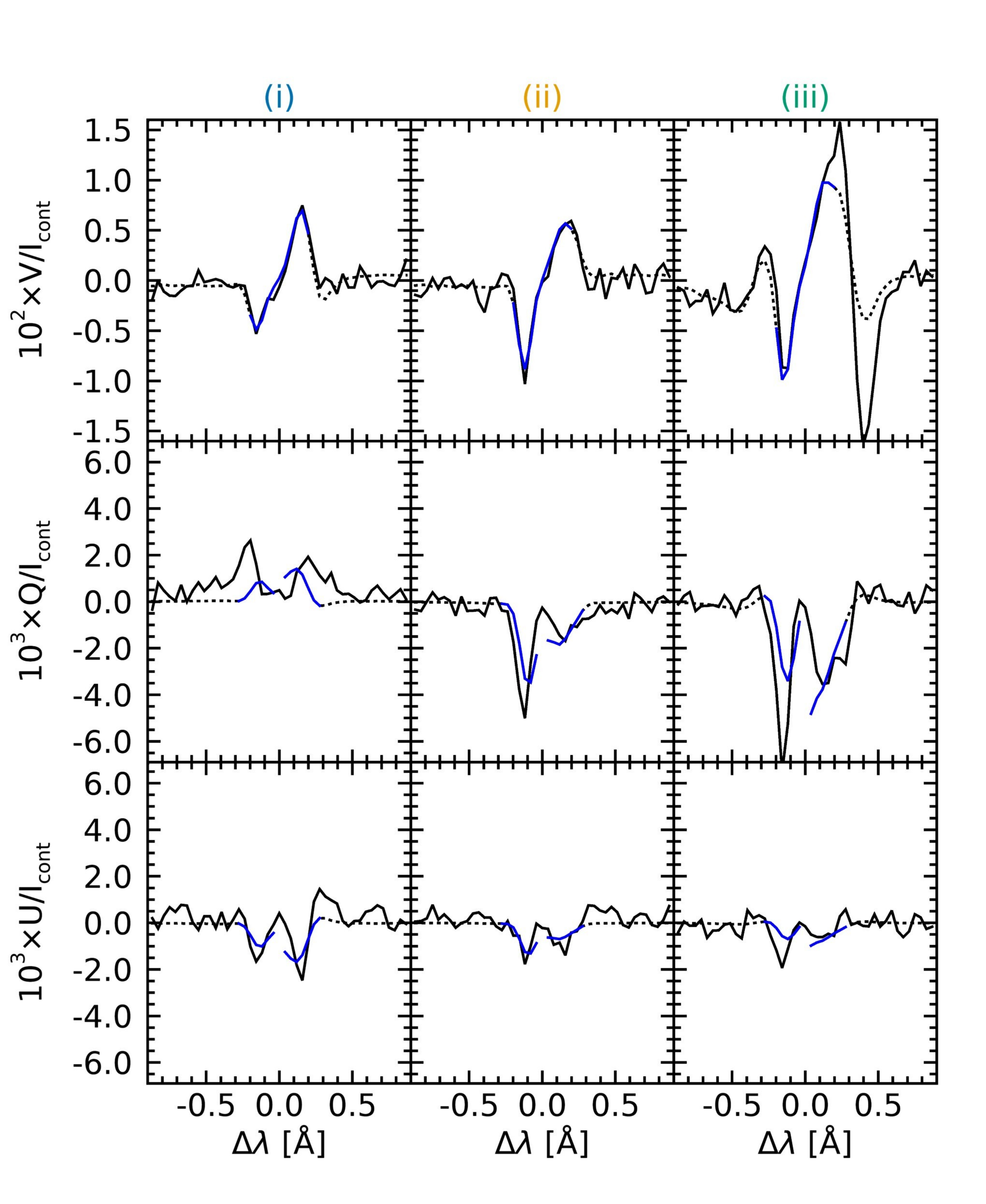}
    \caption{
      Fitting results of WFA at 3 sample points in the filament location, indicated by (i) blue, (ii) orange, and (iii) green crosses in the figure \ref{fig:scanmap}c.
      The black lines indicate the observed Stokes signals, overplotted with the best fit results as the black dotted lines.
      The spectral ranges used for the WFA fit are marked with blue lines.
    }
    \label{fig:wfafit}
\end{figure}

The Stokes spectra observed with \ion{Ca}{2} 8542 {\AA} in the region spatially coinciding with the filament are well reproduced by the Zeeman effect, which generates the characteristic double-lobe structure (effectively a triple-lobe pattern with the central lobe diminished due to line saturation) in the linear polarization (Stokes $Q$ and $U$) profiles, while no significant central single-lobe component is detected \citep{2016ApJ...826L..10S}. 
To infer the magnetic field configuration, we applied the weak field approximation (WFA; \cite{2018ApJ...866...89C,2020A&A...642A.210M}), under the assumption that the Zeeman effect dominates the observed polarization and that Hanle contributions are negligible. 
Figure \ref{fig:wfafit} shows the observed Stokes profiles at three representative positions, marked with (i) blue, (ii) orange, and (iii) green crosses in Figure \ref{fig:scanmap}c. 
The wavelength intervals used for the fitting, $[\lambda_0 - 0.2$ {\AA},$\lambda_0 + 0.2$ {\rm \AA}] for Stokes $V$ and $[\lambda_0 - 0.28$ {\AA}$,\lambda_0 + 0.28$ {\AA}] without line center to avoid zero division for $Q$ and $U$, are marked by blue sector in dotted lines.
The fitting intervals for $Q$ and $U$ were selected to exclude the line center and inner core, while avoiding the far wings where significant photospheric signals would dominate \citep{2018ApJ...866...89C}.
The fitting results for (i), (ii), and (iii) are ($B_{\rm LOS},B_\perp$) = (-86, 364), (-94, 396), and (-122, 458) G, respectively.

\begin{figure*}
	\includegraphics[width=\linewidth]{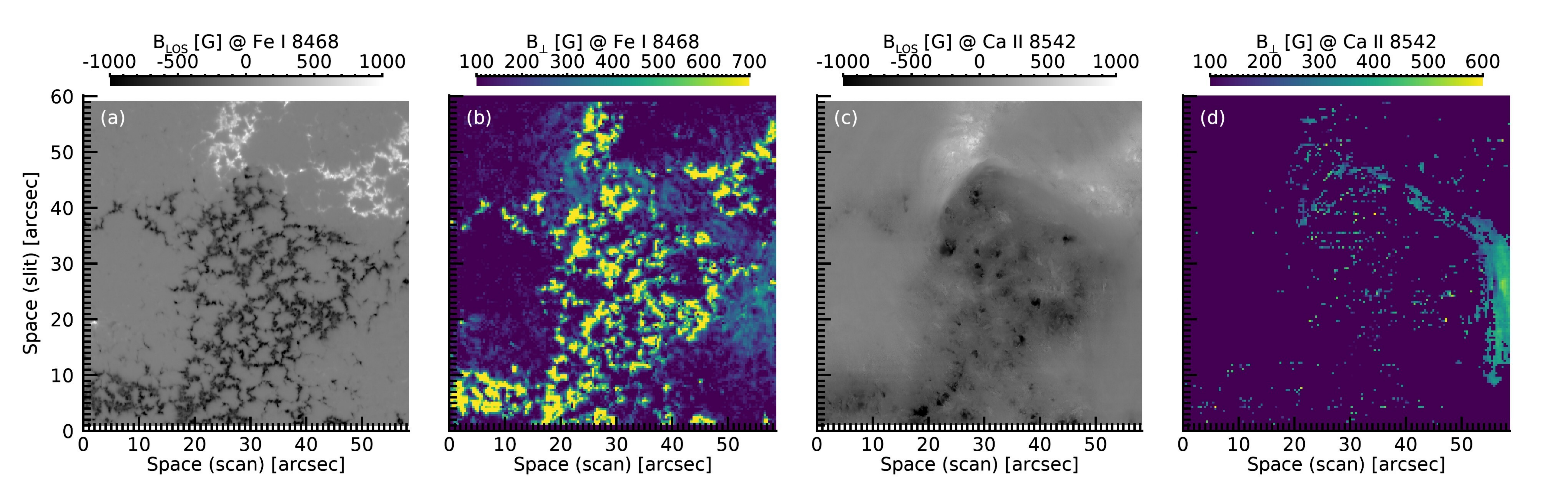}
    \caption{ 
    2D maps of (a) $B_{\rm LOS}$ from \ion{Fe}{1} 8468 {\AA}, (b) $B_{\perp}$ from \ion{Fe}{1} 8468 {\AA}, (c) $B_{\rm LOS}$ from \ion{Ca}{2} 8542 {\AA}, and (d) $B_{\perp}$ from \ion{Ca}{2} 8542 {\AA}.
    }
    \label{fig:magnetogram}
\end{figure*}

Figure \ref{fig:magnetogram} shows the 2D scan maps of (a) $B_{\rm LOS}$ from \ion{Fe}{1} 8468 {\AA}, (b) $B_{\perp}$ from \ion{Fe}{1} 8468 {\AA}, (c) $B_{\rm LOS}$ from \ion{Ca}{2} 8542 {\AA}, and (d) $B_{\perp}$ from \ion{Ca}{2} 8542 {\AA}.
The LOS magnetic field for the \ion{Fe}{1} line was derived using the center-of-gravity method \citep{1979A&A....74....1R,2003ApJ...592.1225U}, while the WFA was applied for the other variables.
The filament lies above the polarity inversion line in the photosphere.
The photospheric LOS magnetic field shows highly localized concentrations that exceed 1 kG.
The photospheric transverse magnetic field is also strong above these LOS concentrations, whereas its strength below the filament is typically around 180 G.
In the chromosphere, the overall LOS magnetic structure broadly resembles that of the photosphere.
The chromospheric transverse magnetic field (Fig \ref{fig:magnetogram}d) above the filament region is also significant, typically 300–500 G, though with larger uncertainties than the LOS component due to the limited signal-to-noise ratio and the applicability limits of the WFA.
These values are substantially larger than the photospheric transverse field (180 G), indicating that the photospheric field alone cannot explain the inferred chromospheric field strength.
Moreover, the bias of the maximum-likelihood estimator for $B_\perp$ \citep{2012MNRAS.419..153M} was 80 to 180 G above the filament region and was well below the measured field strength, confirming that noise in $Q$ and $U$ alone cannot account for the observed field strength.
The derived values represent the integrated magnetic properties along the line of sight within the formation height of the Ca II 8542 line, which may include contributions from both the filament threads and the underlying chromosphere.

\begin{figure*}
	\includegraphics[width=\linewidth]{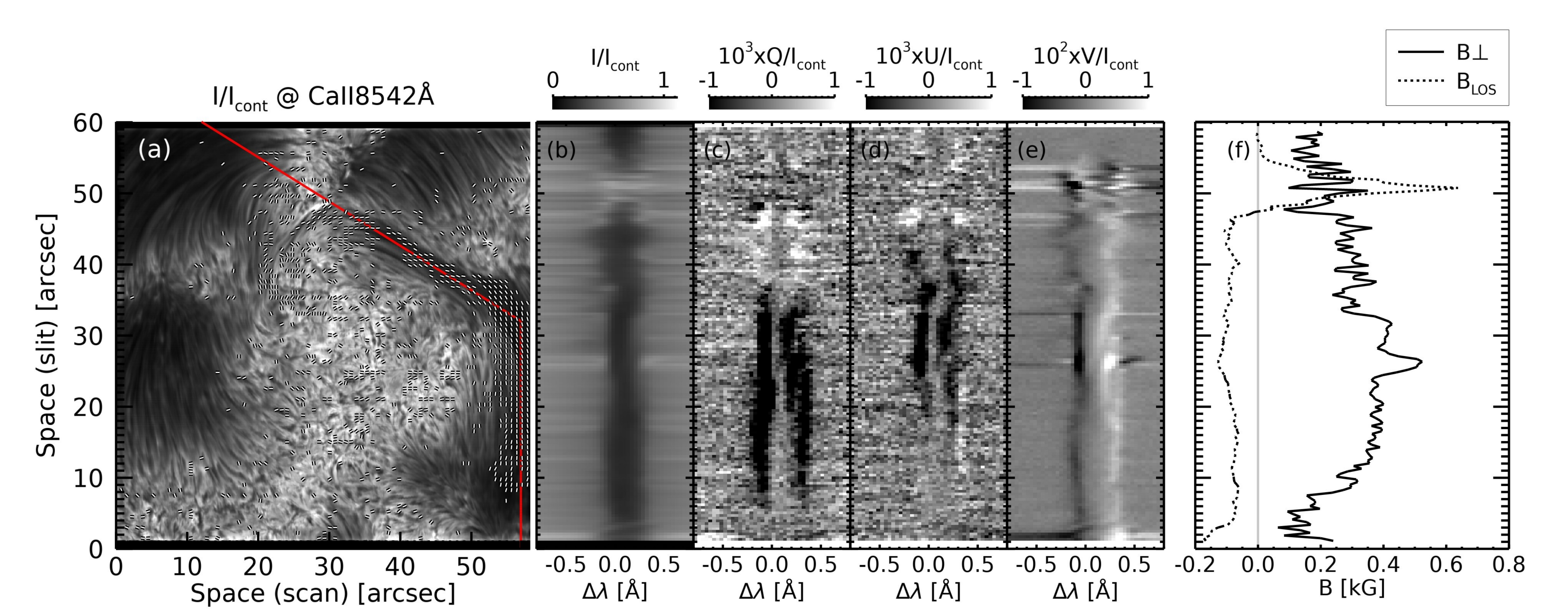}
    \caption{ 
    Stokes profiles along the filament.
    Panel (a) shows the line-core intensity map of \ion{Ca}{2} 8542 {\AA}, the same as in Figure \ref{fig:scanmap}a, overplotted with the azimuth angle of the linear polarization as white line segments.
    The Stokes profiles along the red line in panel (a) are shown in (b) Stokes $I$, (c) $Q$, (d) $U$, and (e) $V$.
    The derived chromospheric magnetic field obtained using the WFA is shown in panel (f), with $B_{\rm LOS}$ as the dotted line and $B_{\perp}$ as the solid line.
    }
    \label{fig:along_filament}
\end{figure*}

Figure \ref{fig:along_filament} shows the Stokes profiles and the derived magnetic field along the filament.
The azimuthal directions of the linear polarization are overplotted as white line segments on the line-core intensity map in panel (a).
Under the WFA, these azimuth angles represent the orientation of the horizontal magnetic field.
In the upper portion of the filament, which is fully contained within the FOV, the magnetic field orientation is nearly aligned with the filament axis.
However, for the lower portion, the filament axis gradually deviates from the FOV (see Fig\ref{fig:fov}c,d).
Therefore, we limit our discussion regarding the alignment of the magnetic field and the filament spine to the upper section.
Panels (b)–(e) present the Stokes $I$, $Q$, $U$, and $V$ profiles along the filament axis indicated by the red solid bent line in panel (a).
The linear polarization profiles exhibit a characteristic double-lobe structure, consistent with the expectation from the transverse Zeeman effect, supporting the interpretation that the measured signals are mainly of Zeeman origin.
Panel (f) displays the derived chromospheric magnetic field along the filament axis: $B_{\rm LOS}$ remains nearly constant at about -80 G, while $B_\perp$ ranges between 300 and 500 G.
The measured LOS field strength of 80 G is consistent with \cite{1991A&A...247..379W} or slightly larger than those measured with previous \ion{Ca}{2} 8542 {\AA} observations \citep{2019A&A...623A.178D,2012SoPh..280...69H}.
The total field strength measured with \ion{He}{1} 10830 {\AA} spans a wide range from 10 to 800 G \citep{2009A&A...501.1113K, 2011A&A...526A..42S,2012ApJ...749..138X,2014A&A...561A..98S,2016ApJ...822...50D,2026PASJ...78..252H}, with our measured values falling within this range.
Notably, \citet{2016ApJ...822...50D} suggested, based on \ion{He}{1} inversions, that the magnetic field in active-region filaments could be as weak as 10 G.
Such a difference in field strength does not necessarily imply a contradiction, but rather highlights that \ion{Ca}{2} 8542 {\AA} diagnostics probe the lower-chromospheric portion of the filament channel, which can be more intensely magnetized than the regions typically observed in \ion{He}{1}.
Recently, \cite{2026PASJ...78..252H} also reported a maximum field strength of at most 120 G for 3 active-region filaments observed at the limb, which is significantly lower than our derived values.
While these discrepancies might be attributed to the intrinsic or height-dependent variability among individual filaments, it should be noted that our results are based on the simplified assumptions of the WFA, and thus a cautious interpretation of these absolute field strengths is warranted.

  In our case, at least in the upper portion of the filament where the full structure is captured within the FOV, the magnetic field is almost parallel to the filament axis within the intrinsic estimation error for WFA of about 10$^\circ$ \citep{2018ApJ...866...89C,2021ApJ...911...23J}, and the line-core intensity map shows thread-like structures inside the filament following the same orientation.
  The observed alignment between the filament axis and the inferred magnetic field orientation in the Ca II 8542 diagnostics may be consistent with either a highly sheared arcade or a weakly twisted flux rope configuration, although the measured signals likely include contributions from both the filament and the surrounding chromosphere.

Although the \ion{Ca}{2} 8542 {\AA} line is one of the most widely observed chromospheric diagnostics, high-precision spectropolarimetric measurements of this line above filaments are remarkably scarce.
Most existing datasets were obtained with imaging spectropolarimeters such as CRISP, which can reach polarimetric sensitivities of order $10^{-3}$ under excellent conditions \citep{2019A&A...623A.178D}.
Spectropolarimeters such as SOLIS \citep{2012SoPh..280...69H}, SPINOR \citep{2014ApJ...788..183B}, and GRIS \citep{2025A&A...698A..33Q} are, in principle, capable of even higher sensitivity.
However, despite these instrumental capabilities, we could not identify any published detections of linear polarization from filaments in the \ion{Ca}{2} 8542 line.
Previous observations with CRISP and SOLIS have reported circular polarization associated with filament magnetic fields, but no studies appear to have detected or analyzed clear $Q$, $U$ signals from filaments in this spectral line.
The reasons for this observational gap remain unclear, but several possibilities exist: filaments may commonly possess relatively weak transverse fields at the \ion{Ca}{2} 8542 formation height, or strong sub-resolution structuring may produce significant cancellation of $Q$ and $U$ within a resolution element.
Such effects would naturally hamper the detectability of linear polarization even when the nominal polarimetric sensitivity is sufficient.
In this light, 
the observed polarization signals may indicate an unusually strong and/or coherent magnetic structure associated with this filament.
The stable, seeing-free observing environment of SCIP likely played a crucial role.
In particular, the absence of atmospheric jitter allowed us to perform spatial binning effectively to lower the noise level while maintaining sufficient spatial information, thereby achieving a polarimetric sensitivity high enough to extract these subtle signals.

The WFA fitting sometimes shows significant residuals.
This is especially true above the cores of the magnetic concentrations and near the very center of the filament (e.g. case iii in Fig. \ref{fig:wfafit}).
The pronounced residuals above the magnetic concentrations and near the filament center likely reflect departures from the assumptions of the WFA, in particular LOS gradients of the magnetic field and velocity and/or unresolved multi-component structures within the resolution element.
In particular, the superposition of the filament threads and the underlying chromospheric plasma along the line of sight may create complex, multi-component Stokes profiles that the WFA cannot fully disentangle.
More sophisticated inversion techniques \citep{2015A&A...577A...7S,2019A&A...623A..74D,2022A&A...660A..37R,2022ApJ...933..145L} that can account for such vertical stratifications and multi-component structures may therefore be required to clearly distinguish the magnetic field of the filament from that of the underlying chromosphere.

\section{Conclusions} \label{sec:concl}

We have presented high-sensitivity spectropolarimetric observations associated with a solar filament obtained with SCIP onboard the \textsc{Sunrise~iii} balloon-borne solar observatory.
The observations targeted an active region filament located near the solar disk center.
Utilizing full-Stokes measurements in the Ca II 8542 {\AA} infrared line, we clearly detected both circular and linear polarization signals associated with the Zeeman effect.

This constitutes, to our knowledge, the first clear detection of linear polarization using \ion{Ca}{2} IR lines at the location of a filament, complementing the traditionally used \ion{He}{1} line.
The polarization profiles exhibit double-lobed signatures in the Stokes $Q$ and $U$ components, characteristic of the Zeeman effect and unambiguously indicative of the presence of horizontal magnetic fields.

The orientation of the linear polarization is found to be almost parallel to the filament axis, and the application of the WFA allows for estimation of the integrated magnetic field strength across the filament and the underlying chromosphere.
These results provide a new observational constraint on filament magnetic structure in the lower chromosphere and demonstrate the powerful capability of SCIP for probing chromospheric magnetism.

Our findings open new possibilities for routine and quantitative measurements of magnetic fields in filaments using high-resolution infrared spectropolarimetry.
Future observations with similar instruments can significantly advance our understanding of the coupling between magnetic field and the chromospheric plasmas.

\begin{acknowledgments}
We extend our heartfelt gratitude to the reviewer for their insightful comments and constructive feedback, which greatly enhanced the quality of this manuscript.
\textsc{Sunrise~iii} is supported by funding from the Max-Planck-Förderstiftung (Max Planck Foundation), NASA under Grant \#80NSSC18K0934 and \#80NSSC24M0024 (“Heliophysics Low Cost Access to Space” program),  and the ISAS/JAXA Small Mission-of-Opportunity program and JSPS KAKENHI Grant Numbers JP18H05234 and /JP23K25916.
This research has received financial support from the European Union’s Horizon 2020 research and innovation program under grant agreement No. 824135 (SOLARNET) and No. 101097844 (WINSUN) from the European Research Council (ERC). It has also been funded by the Deutsches Zentrum für Luft- und Raumfahrt e.V. (DLR, grant no. 50 OO 1608).
The Spanish contributions have been funded by the Spanish MCIN/AEI under projects RTI2018-096886-B-C5, and PID2021-125325OB-C5, and from “Center of Excellence Severo Ochoa”' awards to IAA-CSIC (SEV-2017-0709, CEX2021-001131-S), all co-funded by European REDEF funds, “A way of making Europe''.
The research activities and the flight operation of the SCIP team members, R. T. Ishikawa, M. Kubo, Y. Kawabata, and T. Oba, have been supported by JSPS KAKENHI Grant Numbers JP23KJ0299, JP24K07105, JP23K13152, and JP21K13972, respectively.
Data analysis were partly carried by using the computational resource of the Center for Integrated Data Science, Institute for Space-Earth Environmental Research, Nagoya University through the joint research program.
CQN acknowledges support from Grants PID2022-136563NB-I00/10.13039/501100011033, and PID2024-156538NB-I00 and PID2024-156066OB-C55 funded by MCIN/AEI/10.13039/501100011033.
This work uses the data from CHASE mission supported by China National Space Administration.
\end{acknowledgments}

\bibliography{myref}{}
\bibliographystyle{aasjournalv7}

\end{document}